\documentclass[reprint,
 amsmath,amssymb,
pra,showkeys
]{revtex4-2}

\usepackage{academicons}

\usepackage{url,hyperref,lineno,microtype,subcaption,listings,xcolor,graphicx}
\usepackage{booktabs}

\usepackage{xr}
\usepackage{caption}
\hypersetup{breaklinks=true}
\begin{document}

\title[Measuring Research Information Citizenship Across ORCID Practice]{Measuring Research Information Citizenship Across ORCID Practice}

\author{Simon J Porter \href{https://orcid.org/0000-0002-6151-8423}{https://orcid.org/0000-0002-6151-8423} }%
 \email{s.porter@digital-science.com}
 
\affiliation{%
 Digital Science, 6 Briset Street, London, EC1M 5NR
}%

\begin{abstract}
Over the past 10 years stakeholders across the scholarly communications community have invested significantly not only to increase the adoption of ORCID adoption by researchers, but also to build the the broader infrastructures that are needed both to support ORCID and to benefit from it. These parallel efforts have fostered the emergence of ``research information citizenry'', which comprises, but is not limited to, researchers, publishers, funders, and institutions. This paper takes a scientometric approach to investigating how effectively ORCID roles and responsibilities within this citizenry have been adopted. Focusing specifically on researchers, publishers, and funders, ORCID behaviours are measured against the approximated research world represented by the Dimensions dataset.
\end{abstract}


\maketitle

\section{Introduction}

In 2012, the founding members of ORCID Consortium asked the scholarly community to join them in imagining a new version of the scholarly record: One in which researchers were globally and uniquely identified  \citep{10.1087/20120404}. Although this sounds like a simple, incremental step, it was much more fundamental, at once solving information ambiguities and addressing issues of identity in an increasingly international community where trust in the validity of authorship is a critical currency.  On a practical level, by attaching their ORCID iD to research objects such as publications, researchers would be able to reduce administrative burden of communicating who they are and what they do across multiple  domains including publishing, institutional assessment, research funding, and scholarly information discovery. Institutions within these domains would, in turn, gain greater strategic insight from the scholarly record not readily realisable within their own information silos. 

Even at the beginning of the ORCID project, it was understood that to realise the benefits of ORCID, social and cultural change would be required in addition to technical change. Sustained community investment and collaboration around the development of ORCID and related infrastructures would need to be established amongst a disparate group of stakeholders with different drivers and motivations. All would need to be committed to developing and adopting new workflows and methods of information exchange.  By connecting themselves to, and relying on each other, this newly networked community of researchers, institutions, funding
bodies, publishers, and research service providers would establish the foundations of a new research information citizenship \citep{science_porter_2016}, defined by researcher agency, and distributed metadata stewardship. 

When we speak about researcher agency we think about a combination of the researcher owning their digital representation in the form of an ORCID as well as interacting with the digital world through their representation. By implicitly establishing this concept of researcher agency, ORCID upended passive assumptions about how a research identifier could be deployed. An ORCID iD was not just an identifier for a researcher that could be added by anybody to a record, it was also an identity through which a researcher could interact. In addition to creating trusted assertions within publisher, funder and other administrative workflows, a researcher could also gain access to research services including research facilities and collaboration tools at both an administrative level of securing that access as well as at the practical level of logging into a piece of equipment to perform their work. This merging of the worlds of describing research and conducting it created the possibility that trusted metadata about who was doing what research could be a byproduct of research itself.

Distributed metadata stewardship arises as a natural consequence of researcher agency in a complex ecosystem of stakeholders: It is simply not efficient, desirable or practical to try to centralise activities that researchers are used to transacting with different stakeholders\footnote{The idea of a centralised identity and authentication mechanism for academia is an alluring one.  However, the idea that, at the current time, publishers, funders and academic institutions would all make themselves reliant on a centralised third-party is difficult to imagine.  Furthermore, we live in an era where the direction of movement in technology is toward the decentralisation of trust or, more specifically, the distribution of trust across networks.  Hence, it seems unlikely that centralisation in this context would be a wise structural choice at this time.}. As researchers engage across the activities in the research life cycle, different parts of the data contained in the ORCID registry of scholarly activities are made available to and shared across many different systems. In the case of publishing, a set of authenticated ORCID relationships between a set of researchers and a publication is collected at the time of submission or during the publication process. This distributed authentication is important as capturing these relationships at the point of submission is one of the few times when there is an incentive that can be applied in favour of data quality.  A further consequence of distributed metadata stewardship is that the scholarly record itself becomes distributed, with different stakeholders holding differing levels of detail about each ORCID in their own systems. For instance, a publication identified by a DOI supplied by Crossref records the link between an ORCID iD and a specific author on the paper \citep{clark_2020}, whilst an ORCID record at orcid.org  records the direct connection between a publication and researcher \citep{orcid_2021}.  While the distributed nature of this approach to data holding adds a level of privacy for an individual (since no one actor or system has access to all the information about that individual) there are also pitfalls - specifically, the opportunity for data loss or data inconsistency.  Without a single source of truth or a set of mechanisms to homogenise data (such as a distributed data ledger) there is always the possibility of data ambiguity.

In addition to changes in workflows and responsibilities, global adoption of ORCID has also required a global network of change agents.  Rather than being ``top-down'' initiatives led by governments, the mainstay of these activities has been done slowly with a mixture of bottom-up approaches and mid-level interventions.   Country-led ORCID Consorita have organised to help researchers understand the benefits of maintaining their ORCID record.  For their part, funders and publishers initially made ORCID optional in their grant and publication submission processes.  Although for the last few years this has increasingly moved to  requiring researchers to supply their ORCID as part of these processes \citep{orcid_open_letter}. Some countries have also chosen to act at a higher level and now mandate the use of ORCID iDs as part of their researcher reporting processes \citep{orcid_finland}. 

While nudges and mandates can be powerful in gaining adoption, it is easier to achieve compliance if there is a tangible benefit to researchers and other stakeholders.  In parallel with the development of the technology and compliance landscape, infrastructure has been developed to facilitate these benefits. These changes would not be uniform, with Funders and Publishers moving towards ORCID support at different rates depending on their capacity to change their systems to conform with ORCID best practice \citep{mejias_2020}.

Almost a decade on and the success of ORCID can readily be measured by the number of participants actively engaged with ORCID. In 2018, UNESCO reports that the global researcher population had reached 8.9 Million FTE \citep{UNESCO_2021}. At the end of 2018, there were 5.8 Million live ORCID registrations, 1.4M of whom had recorded at least one work \citep{orcid_2021_Apr}. By July 2021, the number of ORCIDs that had an authenticated relationship with at least one scholarly work had increased to 3.9M. That these numbers are even within the same order of magnitude as the UNESCO figure is a significant achievement. While compelling, what these headline numbers don't indicate is the degree to which behaviour and citizenship around ORCID research information has changed. Gaining an insight into the following questions would provide a better understand of how far research citizenship now extends: Are researchers actively using their ORCID throughout the research process, or does observed behaviour simply reflect a compliance response to mandates? Beyond the ORCID registry itself, how are the responsibilities of distributed metadata stewardship being met? Does behaviour differ between countries and disciplines? How far have publishers changed their practices to accommodate ORCID workflows? What is the quality of ORCID metadata outside of the ORCID registry (particularly in the Crossref registry)? We attempt to answer several of these questions in the current paper.

To address these questions, this paper takes a scientomtric approach and analyses ORCID behaviours with reference to the approximated world of researchers as embodied in the Dimensions database. Although not 100\% accurate for all the reasons that ORCID was created in the first place, Dimensions provides a global set of algorithmically created researcher identities against which ORCID uptake can be measured. Additionally, Dimensions global coverage of publications and grants and the links between them provides a sufficient background dataset against which to conduct the analysis. Section \ref{Methods} of this paper provides a description of the methodology used to link ORCID assertions from both Crossref and ORCID with the Dimensions dataset. Section \ref{Results} provides an analysis of the ORCID behaviours that we are able to observe. Finally, Section \ref{Discussion} reflects on the consequences of these findings.

\section{Methods}
\label{Methods}
Previous analyses of ORCID uptake and usage with respect to ORCID's public data file and publication level integration with metadata from Web of Science \citep{dasler_robin_2017_841777}. In this investigation we have used the combined ORCID statements from both the ORCID \citep{blackburn_etal} and Crossref public files \citep{clark_2021} to examine ORCID-related behaviour in publishing as a whole.

\subsection{Data Integration}

To begin our analysis we needed to create a baseline dataset to facilitate comparisons. We generated this baseline by integrating Crossref and ORCID data with Dimensions \citep{10.3389/frma.2018.00023} so that researchers without ORCID iDs could be identified.  Inclusion of the Dimensions data allows us to access enhanced metadata concerning author affiliations, as well as publisher-level and funder-level information.  Dimensions serves as a convenient intersection between the Crossref and ORCID datasets since the construction of Dimensions is predicated on persistent unique identifiers (PIDs) with information from orcid.org already matched back to  Dimensions, and the Crossref data forming a key part of Dimensions' publications data spine \citep{10.1162/qss_a_00112}.  Data from the Crossref public file can be easily integrated at the author level. ORCID and Crossref data were loaded into Google BigQuery, allowing easy integration with Dimensions data, which is also available as a Bigquery dataset \citep{10.3389/frma.2021.656233}. 

Table \ref{Datasources} provides a breakdown of the fields used in the analysis. Data was analysed along the following axis: Publication, Researcher Affiliation (Country), Publisher, Funder, and Researcher Discipline. Of these, Publisher, Funder, and Researcher Discipline are described in further detail below. 

\begin{table}[ht]
\caption{Data sources and fields used in the analysis.\label{Datasources}}
{\begin{tabular}{lll}
\toprule
        Source    & Entity & Metadata Analysed   \\
\midrule
ORCID      & Researcher  & first name, last name, ORCID, \\
& & date ORCID created  \\
      & Publication & DOI \\\hline
Crossref   & Publication & DOI \\
   & Author & first name, last name, ORICD         \\\hline
Dimensions & Researcher & researcher\_id, ORCID, \\
 & & most recent institutional \\ 
 & & affiliation and country\\
Dimensions & Author &  first name, last name \\
Dimensions & Publisher &  publisher \& journal references \\
Dimensions & Funder &  Links between funders and \\ & & researcher \\
\bottomrule
\end{tabular}}
\end{table}

\subsection{Publications}
Publication data from Crossref was integrated with publication data in Dimensions by matching on doi, first name and surname. Reflecting the differences in metadata schemas, publications in the ORCID registry were not matched at the author level, but instead on ORCID iD and DOI. Publications without Crossref DOIs were also ignored as they did not have bearing on the practices measured in this investigation. 

\subsection{Researchers}
Having matched Publications from Dimensions and Crossref at the author level, the corresponding researcher\_id (Dimensions), and ORCID iD (Crossref) could be associated. This match could only be done after having addressed a data quality issue in the Crossref file (described below).

\subsection{Affilitions}

For this analysis the richer set of information around affiliation data in the ORCID record was not used in favour Dimensions data that provided a consistent method of assigning institutional affiliation across researchers with and without ORCID iDs. The most recent affiliation for a researcher was calculated based on the affiliations associated with their most recent publications and grants. 

\subsection{Researcher Discipline}

To facilitate analysis of ORCID adoption by discipline, a researcher's discipline was defined as the two-digit Field of Research classification \citep{ARC_FOR} in which they most commonly publish \citep{porter_2021}. These classifications were assigned by to publications using an NLP approach, ensuring consistency across a global dataset.

\subsection{Data Quality}
\begin{figure}[bh]
\includegraphics[width=0.49\textwidth]{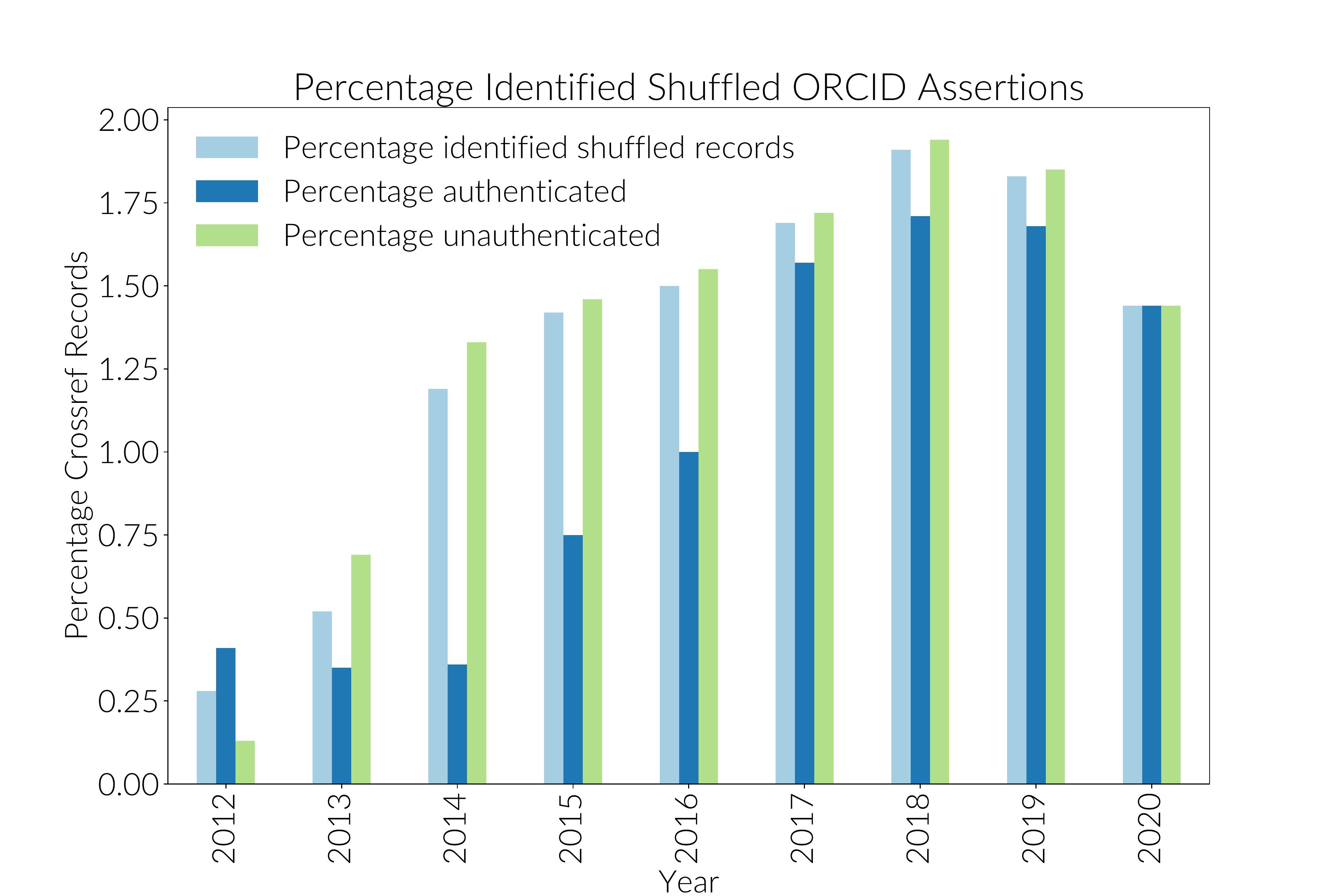}
\caption{Percentage Identified Shuffled ORCID Assertions}\label{fig:suffled-records}
\end{figure}

Before integrating Crossref and ORCID author assertions with Dimensions, Crossref records were first adjusted to address the phenomenon of ``author shuffling''.  (Author shuffling is an effect where by an ORCID iD is assigned to the wrong author on a paper \citep{reflections2021}.)  By joining raw Crossref records to Dimensions records, it was possible to estimate of the size of the author shuffling problem by identifying papers where authors appeared to be collaborating with themselves. In the case of author shuffling, for an author with a reasonably sized publication history, an ORCID iD will be matched to more than one Dimensions researcher\_id. For shuffled records, the research\_id to which they are matched will be one of their collaborators. Shuffled records can be identified when more than one of the researcher\_ids that the ORCID has been associated with appears on the same paper. As Figure \ref{fig:suffled-records} shows, the percentage of shuffled records in Crossref rose to almost 2\% in 2018 before dropping to approximately 1.5\% in 2020. This is almost certainly an underestimate as this method only identifies cases where Dimensions has a researcher\_id for the shuffled author as well as the actual author. The likelihood that a record is shuffled does not appear to be significantly impacted by whether an ORCID iD assertion has been authenticated. This lack of clarity is unsurprising as the data quality issue is systemic.  As of the date of writing this paper, an authenticated ORCID iD statement validates a link between the Researcher and a paper, and does not validate the relationship between an ORCID and a specific author of the paper. 

To increase the chances of finding all shuffled records so that they could be cleaned before matching, suspect author assertions were identified based on the following criteria:
\begin{enumerate}
\item The author appears to be collaborating with themselves (as above), or the match with Crossref results in more than one ORCID iD being assigned to a researcher\_id;
\item The ORCID iD author matched identified by Dimensions disagrees with the author ORCID assertion in Crossref;
\item Dimensions does not have a researcher\_id for the author ORCID assertion in Crossref. 
\end{enumerate}

For these records, a simple string matching algorithm using a Levenshtein Distance calculation was used to establish the most likely match between the name recorded in the ORCID record, and the names of the author on the paper \citep{Cohen2015}. If this approach returned the same match as Crossref with a ratio score of greater than or equal to 70\%, the Crossref match was kept. If the name could be matched to another author on the paper with a confidence score of greater than 90\%, then the ORCID author assertion was reassigned to that author. The difference in confidence cutoffs places a value on the Crossref assertion, as well as addresses a  problem with the matching approach that gave very high scores to incorrectly matched authors with very short first names and surnames. 

One draw back of the above approach to fixing shuffled records, is that it creates a bias against the some of the very use cases that ORCID was established to help solve, including changes in married names, names with few characters, and names with non Latin characters.  In addition, some authors used the native version of their name in their ORCID record, but published with the anglicised version. To help reduce the number of times these instances were rejected due to low name matching scores, author name ORCID matches that could be found across publications from multiple publishers were also accepted as true.  

Using the combination of these methods, 1.7\% connections asserted in the Crossref data were removed, and .5\% reassigned to other authors. That 1.2\% connections were not easily recoverable is illustrative of the difficulty name matching based on strings.


\section{Results}
\label{Results}
\subsection{ORCID Adoption and Engagement}

With the integrated ORCID, Crossref, and Dimensions datasources, we are able to measure ORCID adoption as the percentage of researchers in a given year who have at least one publication with a DOI linked to their ORCID iD either in ORCID directly or identified within the Crossref file. ORCID record completeness was also approximated by comparing the number of publications linked to an ORCID iD vs the number of publications linked to the Dimensions researcher\_id against which the ORCID identifier was matched. As defined, ORCID \textit{adoption} is intended as a measure of active usage,  whereas ORCID record \textit{completeness} is a proxy for engagement. 

We aruge that completeness can be thought of as a proxy for engagement, since a researcher needs to take responsibility for their own record in order for it to be maintained accurately. Firstly, they must  set up their ORCID to receive automatic updates from Crossref, and secondly, they must update their own record with ORCID publication assertions not captured during publisher submission. By including publications in the Crossref record, this measure of completeness is able to include ORCID assertions are not present in a researcher's public record. ORCID assumptions that have been made private by the researcher, and are not included in the Crossref record have not been included in the analysis.

\subsubsection{ORCID Adoption and Engagement by Country}
\begin{figure}[h]
\includegraphics[width=\textwidth]{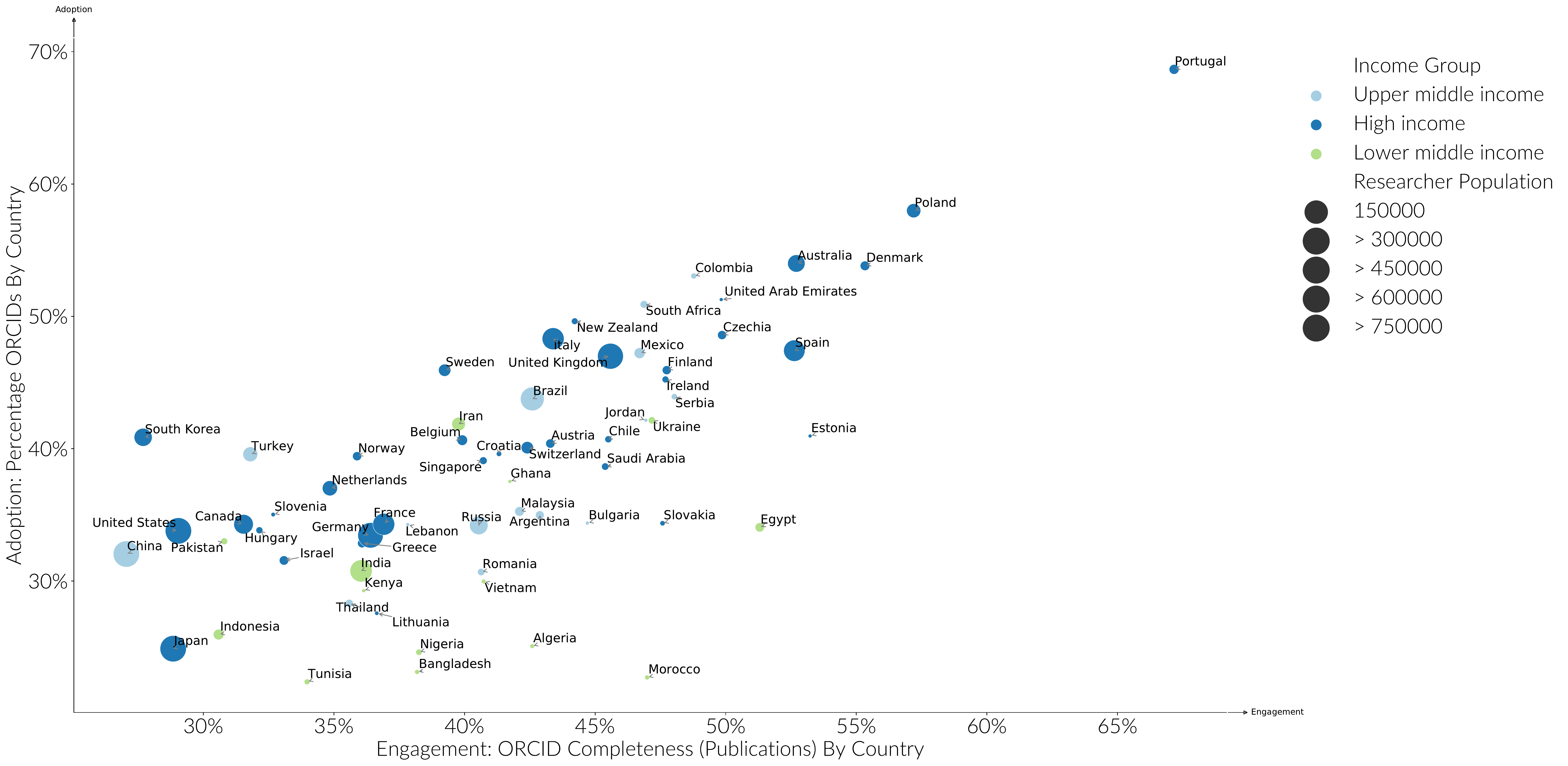}
\caption{Estimated ORCID Adoption and Engagement by Country. Active researchers in the analysis must have a) published between 2015 and 2019, b) have a publication history of greater than 5 years, and c) published more than 5 papers.}\label{fig:ORCID-Adoption-and-Engagement}
\end{figure}

Breaking measures of ORCID adoption and completeness down, by Country (Figure \ref{fig:ORCID-Adoption-and-Engagement} ), it is clear that local research environments significantly influence ORCID researcher engagement. Looking at the years between 2015 and 2019, Portugal ranks most highly in both Adoption (67\%), and Engagement (70\%). Poland, Australia, Denmark, Columbia and South Africa and New Zealand then follow with adoption levels between 50-60\%. Of the countries with an identified researcher pool of $>$ 100,000, the more established and larger scale research economies, Italy, Spain, and the United Kingdom have adoption rates in the region just below or just above  40\%. However, not all the established research economies show the same level of engagement for a cadre of different reasons: The United States, China, and Japan are notable for their relatively low adoption and engagement rates compared to countries of the same world bank income bands  (Table \ref{tab:incomeband}) . In the case of the United States, this is likely to be due to the lack of centralised, government-lead research evaluation and levers associated with block funding the other countries such as those mentioned have available.  Japan has adopted it's own system of researcher identification with the researchmap.jp system, which stands apart from all other global systems.  China, while moving quickly, is simply at an earlier stage of engagement with globalised research infrastructure and has unique challenges in terms of name disambiguation.

\begin{table}
\caption{ORCID adoption and completeness by World Bank income band. Columns: Res. is the number of researchers in the income band; \% is the ORCID coverage for the cohort; Med. gives the median percentage per country in the income group; Comp. gives the percentage completeness of the records for the cohort.}
\label{tab:incomeband}
\begin{tabular}{p{0.14\textwidth}p{0.08\textwidth}p{0.07\textwidth}p{0.07\textwidth}p{0.07\textwidth}}
\toprule
        Income Group & Res. & \% &  Median & Comp. \\
\midrule
 High & 2719500 &  38.52 & 40.52 &  38.14 \\
 Upper middle & 993628 &  36.05 &   32.04 &    33.78 \\
 Lower middle &  361944 &   31.60 &    26.33 &   37.82 \\
 Low &   10274 & 26.70 &   22.22 &   34.45 \\
\bottomrule
\end{tabular}
\end{table}

\onecolumngrid

\clearpage

\onecolumngrid

\begin{figure}
\includegraphics[width=\textwidth]{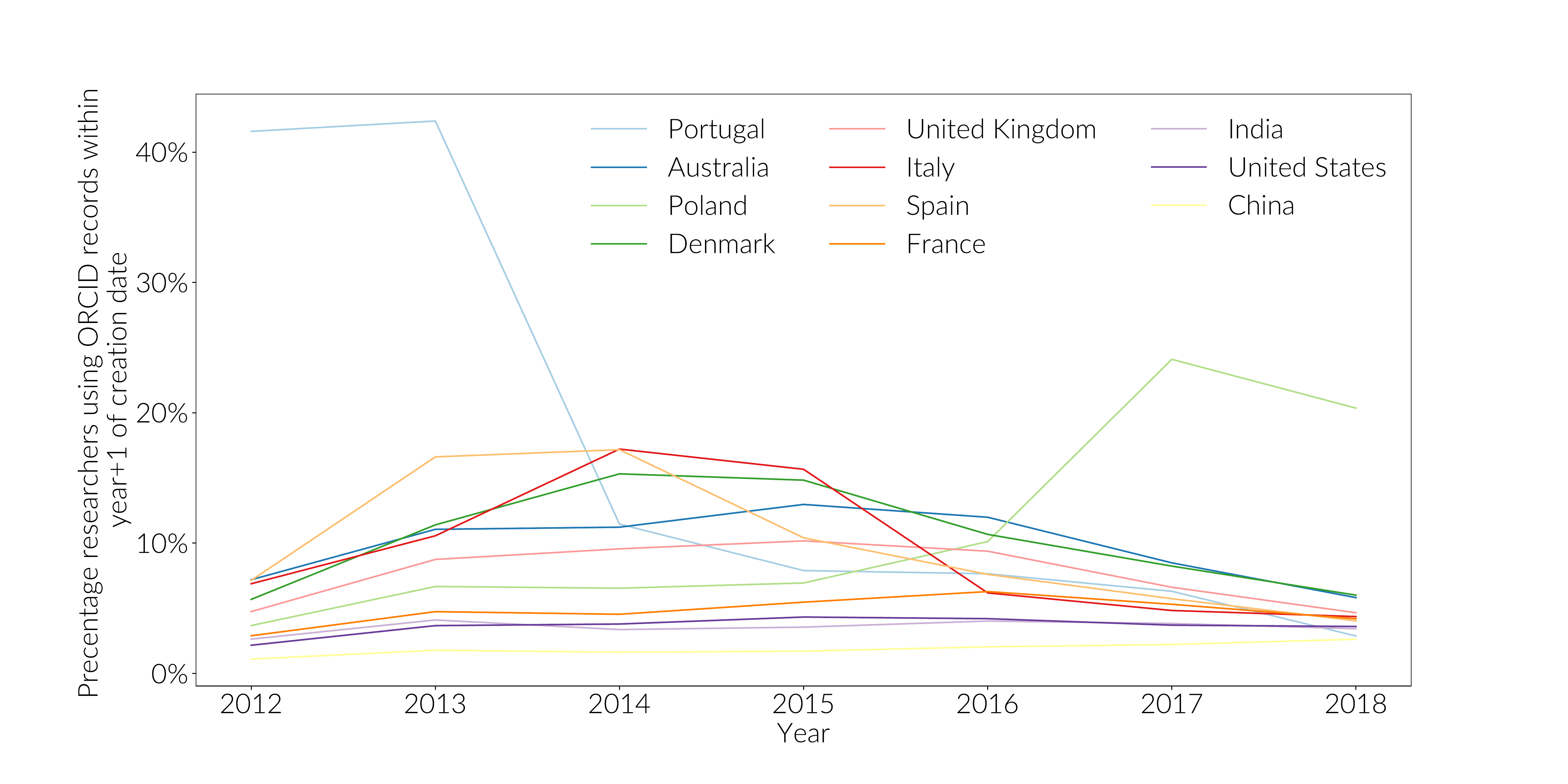}
\caption{New ORCID Registrations by year(+1). Totals are not cumulative, showing early peaks in adoption}\label{fig:New-ORCIDs-by-year}

\end{figure}
\twocolumngrid

Countries with high engagement, have also demonstrated concerted enrolment efforts. These efforts can be detected in the publication record by looking for ORCID iDs that are used in publications between the time they were created and the end of the next full publication year (Figure \ref{fig:New-ORCIDs-by-year}.) Using this methodology, it is possible to observe that Portugal started early with a concentrated effort in 2012, and 2013 at the launch of the ORCID initiative, with Spain following over 2013 and 2014, Italy and Denmark  in 2014-2015. Both Australia and the United Kingdom showed a sustained engagement at or slightly below 10\% between 2013 to 2016. Poland is distinct in initiating renewed engagement activities in 2016.

\begin{figure}
\includegraphics[width=\textwidth]{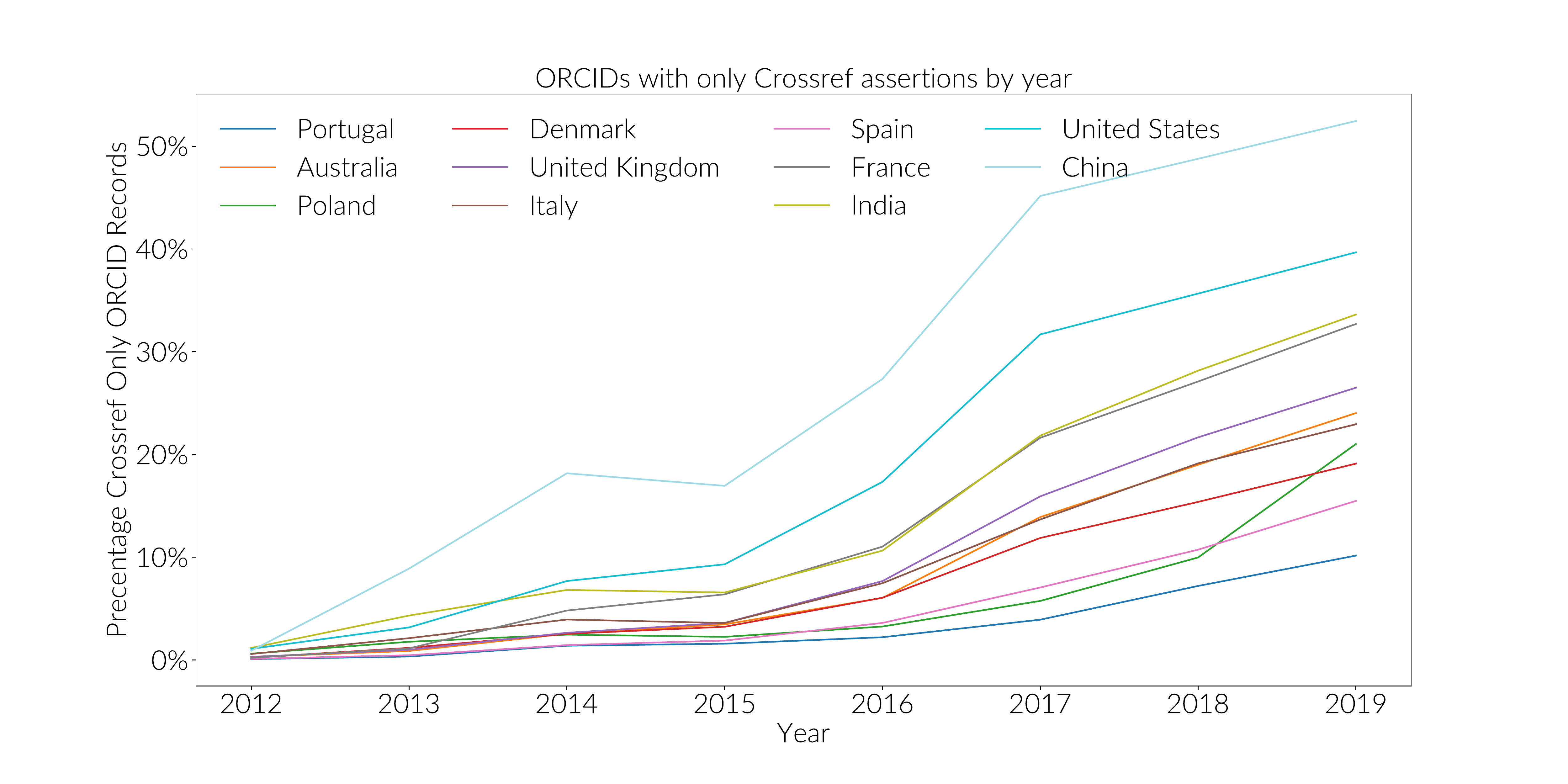}
\caption{Percentage of ORCID iDs with only Crossref assertions by year}\label{fig:ORCID-Crossref-Only-ORCID}

\end{figure}

Countries with low engagement show a different pattern (Figure \ref{fig:ORCID-Crossref-Only-ORCID}). Since 2016, there has been a steep increase in ORCID iD assertions that are present in Crossref, but are not displayed in a researcher's own ORCID record. This is particularly prominent with Chinese authors where 50\% of researchers in 2019 do not have any 2019 statements from Crossref that have made it back to their public  ORICD record. For United States authors, this value 40\%, compared to 10\% for Portugal, and just above 20\% for Italy and Australia. This result is despite the fact that there is an established workflow to push ORCID assertions back from Crossref to  ORCID, and that all researchers are required to do is to provide consent in response to an email \citep{10.3390/publications4040030}. At least two scenarios might explain this behaviour with the strength of this effect varying by country:

\begin{enumerate}
\item  An increasing number of researchers are registering for an ORCID iD because they are encouraged to during early career studies or because need one to engage in certain formal processes within their country, but are not sufficiently engaged to go further and keep their ORCID record up to date, either by entering in details directly or authorising the systems that they engage with to update their record on their behalf such as the Crossref auto update functionality \citep{mejias_2020}
\item An increasing number of researchers are choosing to keep their record private due to growing privacy concerns associated with digital existence as a whole.
\end{enumerate}

The first scenario is concerning as it suggests that a growing number of researchers will not be able to use their ORCID iD as a tool to reduce academic burden. These Researchers will likely be frustrated when the act of supplying their ORCID iD in a funder workflow does not result in their record being populated. This scenario is reasonably likely. In 2017, after the initial release of the Crossref auto-update functionality, only 50\% of researchers were reported as choosing respond to the email from Crossref offering to auto update their ORCID record when new publications were detected \citep{meadows_haak_2017}. For some countries, It does not appear as if this number has significantly improved since this time. 

The second scenario, although not necessarily preventing any ORCID use cases, would indicate an increasing desire by researchers no to be `known' by their ORCID iD, and perhaps a lack of buy in to open identifier infrastructure. Both scenarios would be regional examples of less than enthusiastic research information citizens.

Part of difference between country cultures can be explained by the interventions local funding agencies have made in integrating ORCID iDs into their processes. Funding agencies can impact ORCID behaviour by requiring researchers to have an ORCID (adoption,) as well as by driving engagement by making it easy for researchers to use information from their ORCID records in their publications, or implying a strong preference for complete ORCID records. Beyond publication workflows, funders will also play an increasing role in linking ORICD iDs to open public records of grants \citep{group_2019} creating similar data reuse patterns to publications.

\begin{figure}[h]
\includegraphics[width=.9\textwidth]{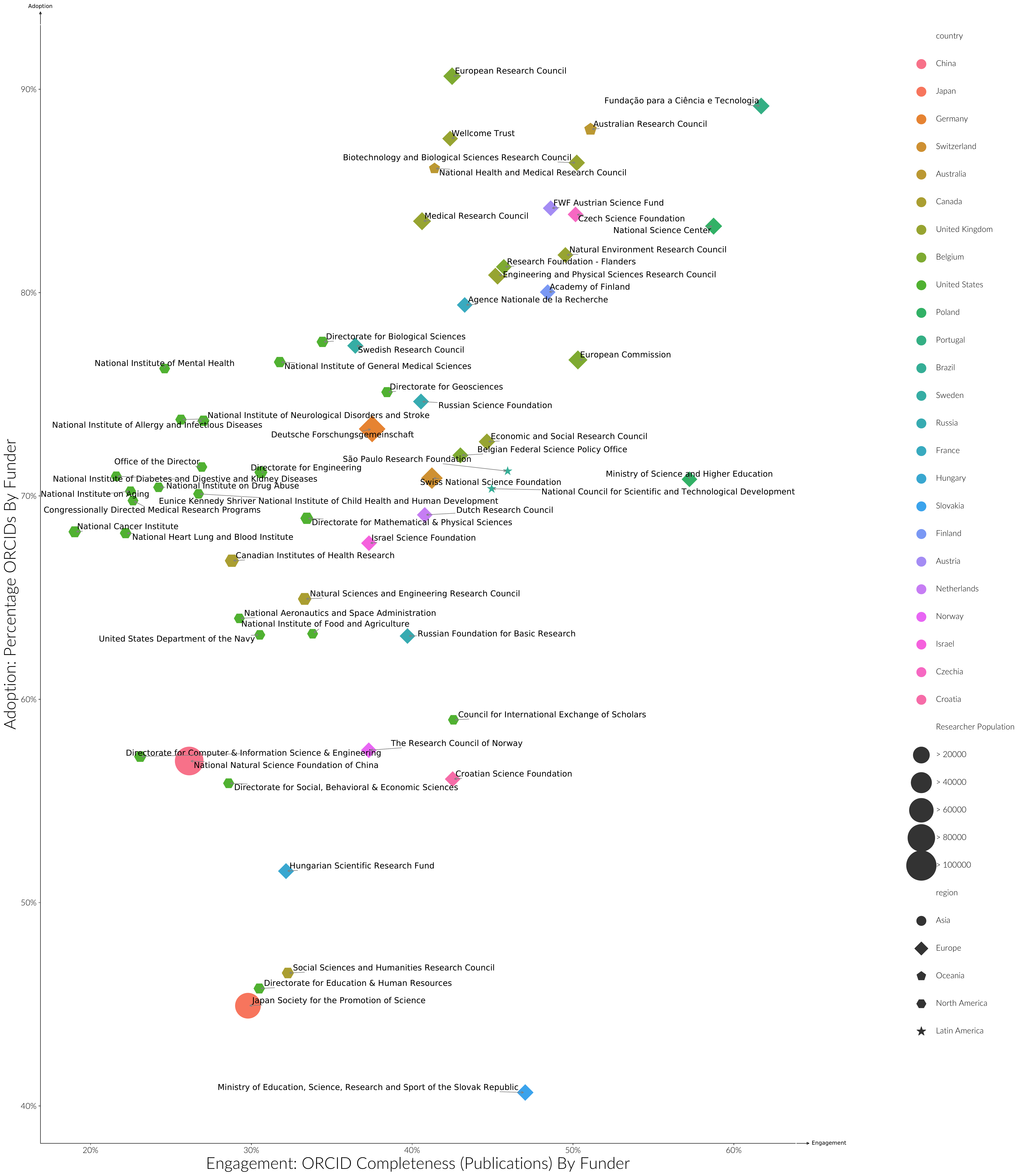}
\caption{Estimated ORCID Adoption and Engagement by Funder
 Researchers >5 years publication history, Publications (2015-2019). Sizes indicate the number of researchers by funder, with shapes denoting world regions.}\label{fig:ORCID-Adoption-and-Engagement-Funder}
\end{figure}

Figure \ref{fig:ORCID-Adoption-and-Engagement-Funder} shows the top 60 funders by the number of researchers with ORCID iDs that they have funded between 2015 and 2019. Across these 60 funders, a much higher ORCID adoption rate can be observed for funded researchers than compared with country averages.  This is to be expected to some degree, as there will be a greater overlap between researchers that receive funding, and researchers required to have an ORCID iD as part of publisher ORCID policies. A similar shift is not observed in the engagement rates by funder when compared to overall country rates.

Even with the overall increase in ORCID adoption rates, distinct funder patterns can be observed. The United Kingdom, Finland, Portugal, Australia, Austria and Czechia have very high adoption rates (between 80\% and 90\%). Many of these funders are associated with funder ORCID policies that either mandate, or strongly recommend the use of ORCID iDs in funder submissions. That engagement rates for these funders do not differ significantly from country norms, suggests an impact beyond just those who were funded, to applicants and the broader community. An underlying information systems capacity for a country to accept a funder mandate may also be in play, with the United Kingdom, Australia, Finland and Portugal, and Czechia all having strong research reporting practices at the country and institutional level. High levels of research engagement implies a high level of ORCID record maintenance. Countries with a mature network of Institutional Current Research Information Systems will be better supported with theses maintenance activities than countries without. 

A separate band of funders including funders from the United States, Canada, Germany, Russia and Israel sees adoption rates between 60\% to 80\%.  Within this second band, where identifiable in funder policies listed by ORCID \citep{orcid_2020_Aug}, ORICD integration funder appears to be more technical and optional rather than policy driven. Other funders within this band have more recently launched ORCID initiatives, the effects of which would not be seen in the analysed period. 

\onecolumngrid

\clearpage

\twocolumngrid

\subsection{ORCID Adoption by Research Category} 
\begin{figure}
\includegraphics[width=\textwidth]{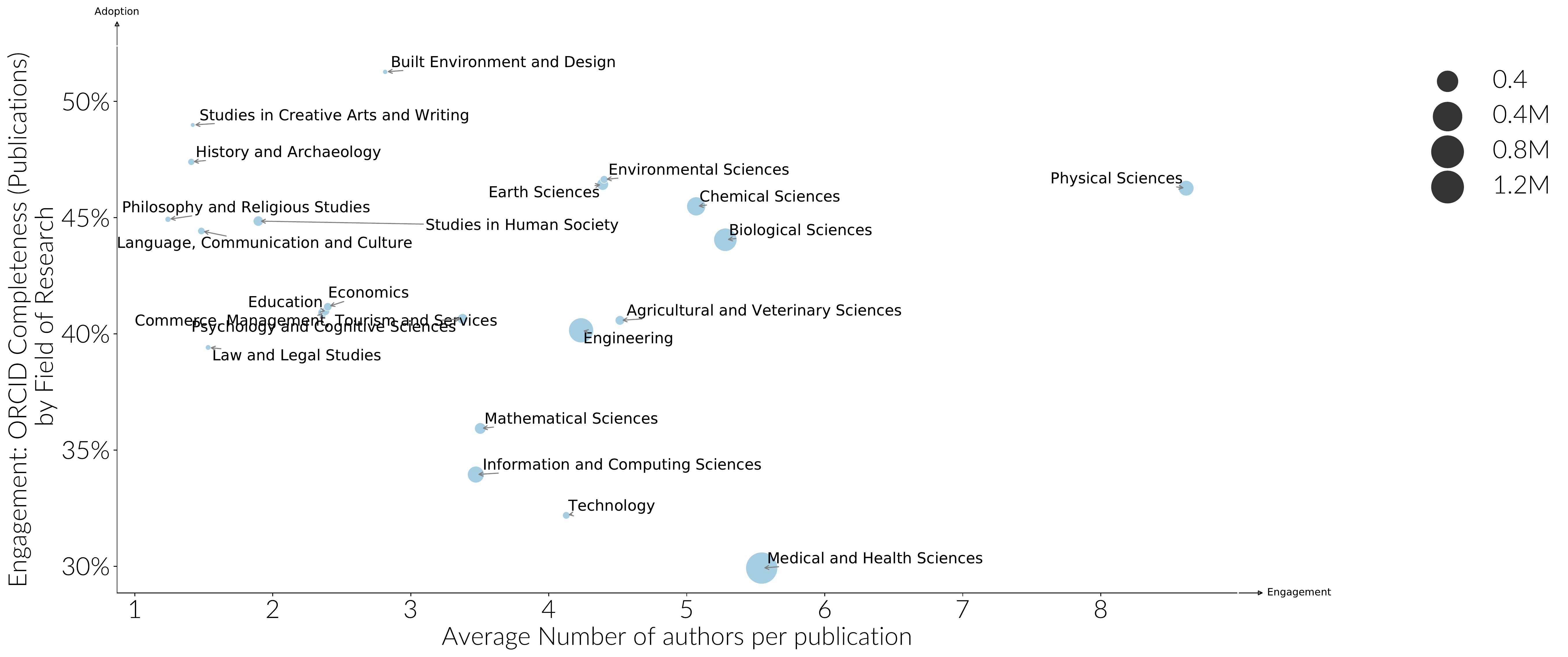}
\caption{Estimated Engagement by Field of Research Compared to the Average Number of Authors per Paper. Researchers are included if they have >5 years publication history, Publications (2015-2019). The size of the marker indicates the size of the identified researcher cohort}\label{fig:ORCID-Classification-avg-authors}

\end{figure}

\begin{figure}
\includegraphics[width=\textwidth]{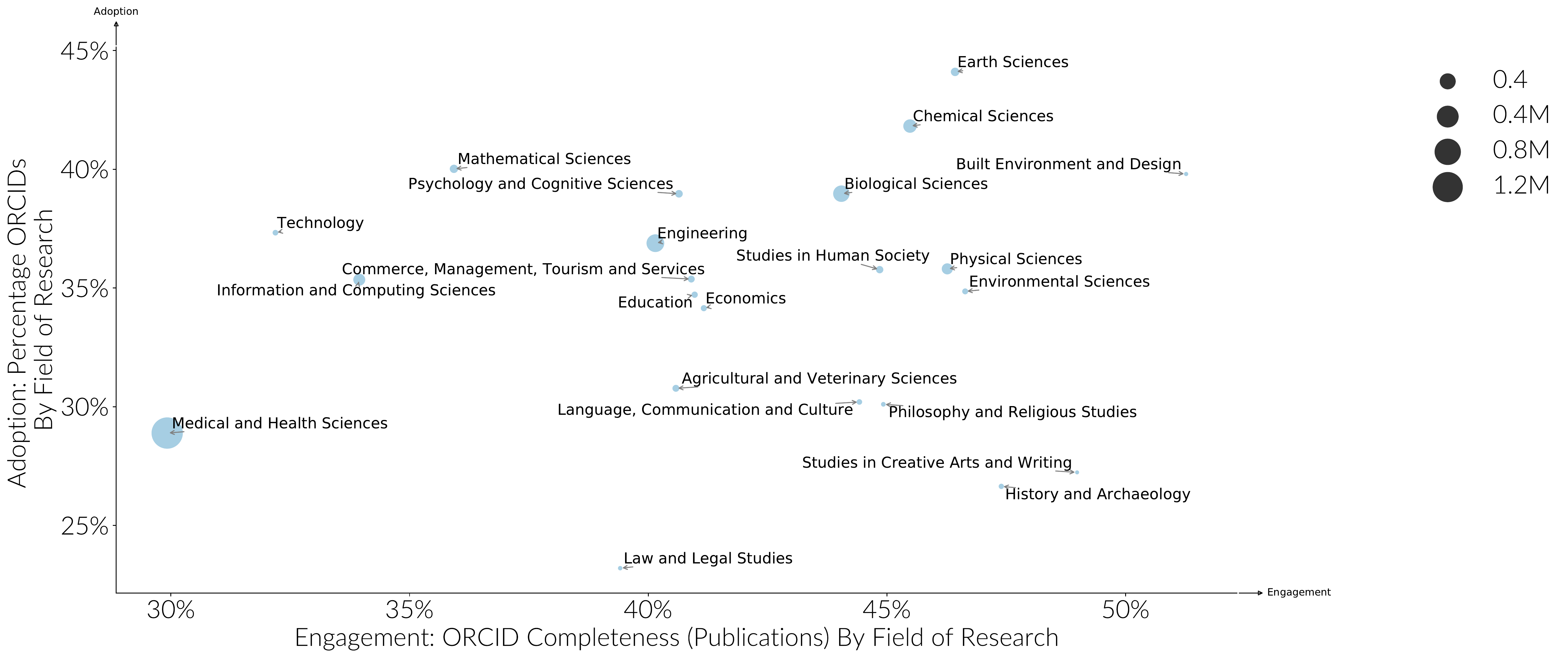}
\caption{Estimated ORCID Adoption and Engagement by Field of Research
 Researchers $\>$5 years publication history, Publications (2015-2019). The size of the circle represents the size of the identified research population}\label{fig:ORCID-Classification}

\end{figure}

Overall, funders adoption and engagement rates are clustered more by country than they are discipline, however some discipline effects can still be observed. Medically focused funders in particular have relatively lower engagement rates on average when compared to other funders in the same country. These differences in discipline are also born out more generally. As shown in Figure \ref{fig:ORCID-Classification}, ORCID adoption by discipline ranges 25\% to 45\%, and engagement from 30\% to 50\%.  Earth Sciences and Chemical Sciences have both high adoption and engagement rates. Humanities research areas are distinguished by having lower adoption levels, but higher engagement levels. The large difference between adoption and engagement levels for these fields is partly explained by the articles in these fields having fewer authors per paper, and therefore fewer middle authors that are unlikely to receive ORCIDs given current publishing workflows. The average number of authors per paper does not explain the disparity in engagement across all disciplines however. For instance, researchers in Medical and Health Sciences have a much lower engagement rate when compared to the relatively high adoption and engagement rates of disciplines with a similar average number of authors per paper such as Chemical or Biological Sciences (Figure \ref{fig:ORCID-Classification-avg-authors}).

As disciplines cross different country and funder environments, a high engagement and adoption levels by discipline suggests that there are pockets of research practice the are closer to normalising the use of ORCID for all authors. 

\onecolumngrid

\twocolumngrid
\clearpage

\subsection{ORCID Adoption - Publisher Level}

Like Funders, Publishers support ORCID adoption and engagement via different mechanisms. ORCID adoption can be driven by publisher mandates, whilst engagement is supported most fully by providing all authors on a paper the opportunity to assert their ORCID iD. Publishers complete their responsibility as research information citizens by passing the ORCID metadata through to Crossref.

\begin{figure}
\includegraphics[width=.85\textwidth]{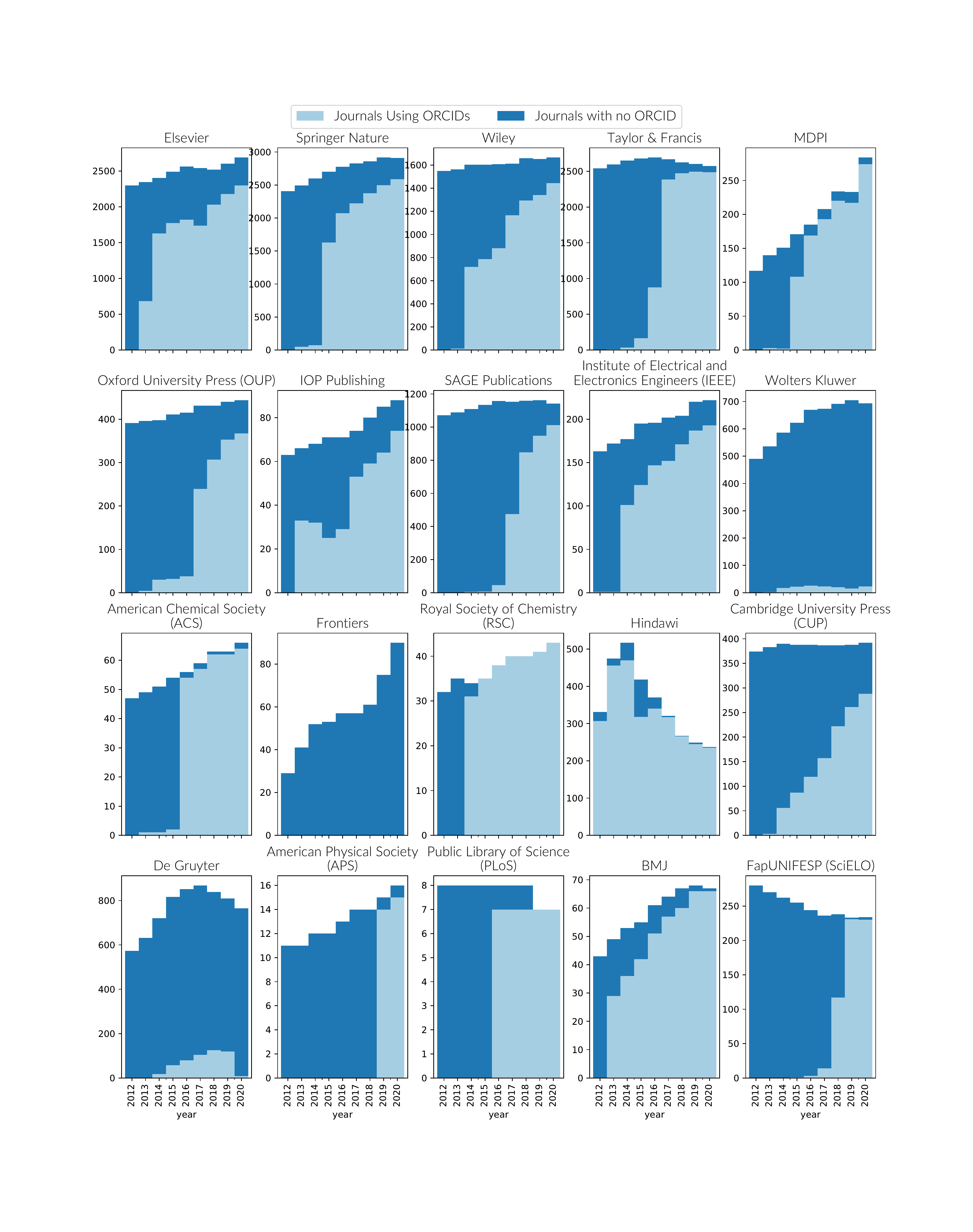}
\caption{Journals supporting the use of ORCID's within Crossref metadata by publisher by year.
 Source: Crossref public file matched to Dimensions}\label{fig:journal-oricd-uptake-by-publisher}

\end{figure}

With a few notable exceptions, support for ORCID in publication metadata by journal and publisher has increased significantly, particularly since 2016. For the top 16 publishers by volume, Figure \ref{fig:journal-oricd-uptake-by-publisher} outlines the number of journals per publisher that have evidence of ORCID metatdata support in their Crossref records. With the exception of Wolters Kluwer, De Gruyter, and Frontiers, near complete journal support for expressing at least a minimum amount of  ORCID metadata has either been reached, or there is a clear trend towards it. Presence of ORCID metadata in the Crossref records is not only a measure for publisher support adoption of ORCID, it is also a measure of community participation in open metadata that can be further consumed by downstream systems - a commitment outlined in the ORCID Open letter for publishers \citep{orcid_open_letter}. In the case of Frontiers, collection of ORCID iDs is a part of their workflow processes, however there was an oversight in passing the information across to Crossref [Internal Communication].

The level of support for ORCID iDs within publications by publisher is less uniform. In 2016, many publishers signed up to the commitment to require at least the corresponding author to connect their ORCID iD, with the understanding that all authors should be provided the option to assert their relationship to the paper \citep{orcid_open_letter}. Most publishers began their implementations by implementing the first requirement  with support for additional authors proceeding at different paces \citep{meadows_haak_2017}. 

\onecolumngrid

\twocolumngrid

\onecolumngrid

\begin{figure}
\includegraphics[width=\textwidth]{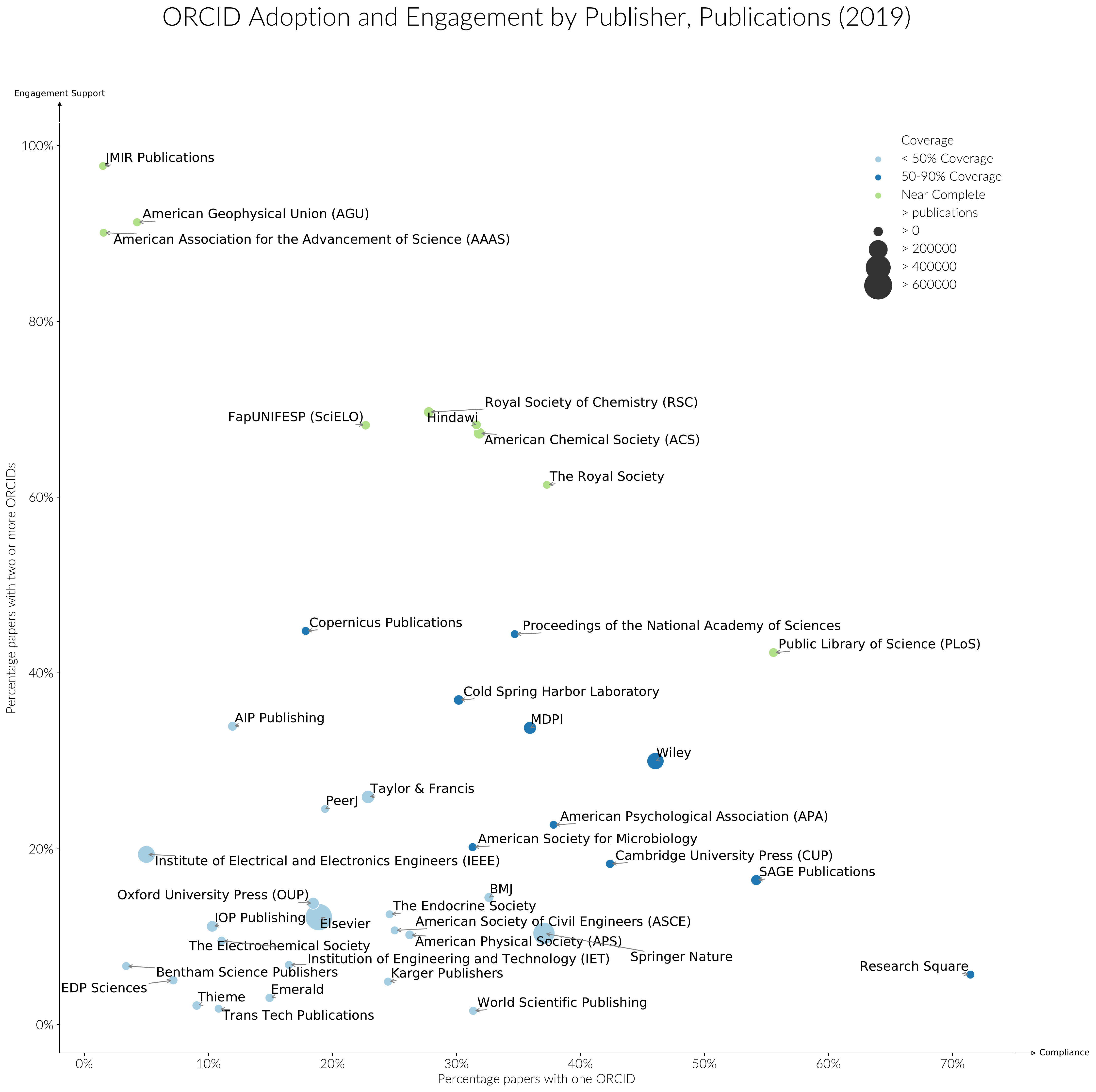}
\caption{ORCID Adoption and Engagement by Publisher, Publications (2019)}\label{fig:ORCID-orcids-per-publisher}

\end{figure}
\twocolumngrid

By looking at papers published in 2019 with more than three authors, it is possible to observe how this trend has since moved. Examining the top 20 publishers by volume of ORCID assertions in 2019 (see Figure \ref{fig:ORCID-orcids-per-publisher}),  the dominant publishing mode was still one ORCID iD per paper, however, clear differences in publishing practice can be observed. Nine publishers had at least one ORCID on over 90\% of their publications in 2019. Of these JMIR, stands out both in the fact that it hast the highest percentage of papers with two or more ORCID iDs, and that its overall discipline that it serves (Medicine and Health Services) does not have a high researcher engagement rate. Eight publishers had a percentage of greater than 60\% papers with two or more  ORCID iDs per paper, with a further band of 7 between 20\% and 40\%. Elsevier and Springer Nature, the largest of the publishers have approximately 10\% of their papers with two or more ORCID iDs, although their coverage of papers with one ORCID iD differs significantly at 18\% and 38\% respectively. That there is such a difference in the spread of support of more than ORCID suggests that the constraint still lies within individual publishing platform implementations, rather than a willingness for researchers to change behaviour.

\section{Discussion}
\label{Discussion}

In the previous section, a scientometric analysis of ORCID behaviours reveals a research information citizenry that are serious about their obligations to each other, albeit one still in transition to ORCID-centric workflows.

We have shown that:
\begin{itemize}
\item In contrast to the internationalisation of research, ORCID adoption and engagement patterns are regional, with countries such as Portugal, Poland, Denmark, and Australia leading the way and research giants such as the United States, China and Japan falling behind.  Researchers within countries with low ORCID adoption rates are also more likely to be disengaged with their profile.
\item ORCID adoption rates for funded researchers are significantly higher than their country averages, reflecting the influence of both publisher and funder mandates
\item Publisher mandates have played a key role in encouraging ORCID adoption, however the capacity for Researchers to supply ORCID iDs is now out significantly outstripping publisher ability to record them as part of the submission process
\item Publishers are meeting their responsibilities for distributed metadata stewardship around ORCID, however there remain some challenges in retrofitting new ORCID processes to existing submission workflows. These challenges resulting in an error rate of ORCID to author assertions of about 1.5\% in 2020. Continued data quality monitoring is essential to ensure that this error rat continues to fall.
\item ORCID adoption and engagement profiles differ significantly by research discipline, with Chemical Sciences and Earth Sciences having the highest rates, and Medical and Health Sciences the lowest. Moving beyond mandates, Innovation in ORCID engagement by discipline provides a sustainable path for ORCID adoption going forward.
\end{itemize}

\subsection{Addressing Researcher Disengagement}

Critically, as might be expected ORCID's success looks different by region, funding regime, and subject area. Each of these factors plays intimately with the likelihood of success of ORCID for a given researcher.  If the researcher works in an established research economy in a high-income country with a dual-funding structure and national evluation in a STEM research area then they are most likely to have both drivers to use ORCID and the opportunity to benefit from infrastructure investments. All this to say that depending on where in the world a researcher is based, they will likely have a significant difference on how integral is ORCID to their daily workflows. 

For ORCID, Research Information Citizenship is not just about having an ORCID iD, but using it in expected ways. For a researcher, a key responsibility is not only ensuring that their information is kept up to date, it is also about ensuring that information can flow into their ORCID record with as little latency as possible.  That countries with low engagement and adoption rates also exhibit a higher rate of disconnection between Crossref and ORCID is of significant concern. As publisher support for ORCID increases, these Researchers are likely to experience the administrative burden of ORCID (which typically impact article submission workflows), without benefiting from the administrative benefits (which typically accrue during national evaluation or funding applications). Strategic engagement of these researchers will not only increase the local benefits of ORCID to the researchers involved, it also offers a path towards reducing the number of 'empty' ORCID profiles.

\subsection{Emerging strains on distributed metadata stewardship}

On the other side of the relationship, it is remarkable that most publishers still publish more publications with only a single ORCID rather than multiple ORCIDs. Pressure to support ORCID assertions for all authors on a publication is mounting, with the capacity for Researchers to supply their ORCID at the time of submission now outstripping functionality to support it. 

Some journals are now choosing to implement ORCID policies that are beyond the current capacity of their publishing workflows \cite{10.1186/s13321-019-0365-4}. For these journals, ORCID iDs will be supplied as part of the submission, however they will be unauthenticated by the researcher themselves leaving open the possibility that a Researcher could be misidentified. It is possible that initiatives designed to increase ORCID engagement could also break community trust by introducing errors into the system.

This pressure on publishers will increase still further with an evolution of funder requirements around open access publishing. UKRI now require all authors to be uniquely identified by their ORCID iD on papers published after April 2022 \citep{ukri_2021}. Notably, the policy does not specifically require ORCID iDs to be authenticated, raising the risk that the number of unauthenticated ORCIDS will rise significantly. This level of funder activism is interesting in that it imposes a mandate on coauthors from other countries to add their authenticated ORCID to UKRI funded publications. 

Overall, Publishers can be seen to be meeting their research information citizenship obligations of passing on metadata through to Crossref.  Although the problem of author shuffling as identified in section \ref{Methods} reflects an inherent difficulty these workflows. At the core of the issue is the task of assigning ORCIDs to the individual author statements made through the manuscript submission process.  Workflows that begin with the free text author statement on a manuscript and require a decision to be made on which author belongs to which ORCID. These decisions introduce name matching errors that are difficult to completely overcome, particularly when retrofitting ORCID to fit over legacy submission workflows. Continued monitoring of author shuffling with feedback to publishers to correct them should be considered an important activity to continue to maintain trust in the ORCID ecosystem.

\subsection{Reflections on the role of scientometric monitoring of ORCID practices going forward}

As demonstrated in this article, scientometric monitoring of ORCID adoption and usage can offer insight in where next to invest in ORCID innovation activities. 

Publisher support and innovation around ORCID may only just be beginning. Within disciplines where ORCID adoption and engagement levels are already high, it might also be possible to turn the relationship between author and ORCID on its head by adopting ORCID first approach to author assertions. Beginning with an ordered set of ORCID iDs, it would then be possible to to derive the authorship statements on a paper.  ORCID iDs could then be authenticated as part of the submission process (or as part of the document authoring process) without the additional requirement for author statement matching. Uncoupling ORCID author assertions from the submission process would also open up opportunities for greater collaboration between publishers and research authoring tools. 

Finally, whilst this analysis has only measured funder contributions to ORCID adoption and engagement rates indirectly, funder interventions can be seen to correlate with high ORCID and engagement rates - particularly amongst countries with well established networks of current research information systems. Until recently, funders have expressed their role a a researcher information citizen as a consumer of ORCID information. More recently \citep{group_2019}, in 2019 an analogous more to the Publisher open letter \citep{orcid_open_letter} a consortia of funders has proposed extending their role to also be a creator of ORCID assertions for grants, by creating both a public record of the grant with a DOI, and an ORCID assertion to it. As these new information pathways establish, scientometric approaches such as the one showcased here will provide an important methodology for charting its progress.

\section*{Conflict of Interest Statement}
The author is an employee of Digital Science, the creator, owner and distributor of Dimensions.

\section*{Author Contributions}

All the work contained in this article was conceptualised, carried out and written up by Simon J Porter.

\section*{Funding}
The author did not receive funding for this paper.

\section*{Data Availability Statement}
The code for grenerating the datasets for this study can be found in figshare: 10.6084/m9.figshare.16638337.

\Urlmuskip=0mu plus 1mu\relax
\bibliographystyle{plain}
\bibliography{bibliography}

\begin{thebibliography}{10}

\bibitem{reflections2021}
Miriam Baglioni, Andrea Mannocci, Paolo Manghi, Claudio Atzori, Alessia Bardi,
  and Sandro~La Bruzzo.
\newblock Reflections on the misuses of orcid ids117-125 miriam baglioni,
  andrea mannocci, paolo manghi, claudio atzori, alessia bardi, sandro la
  bruzzo.
\newblock In Dennis Dosso, Stefano Ferilli, Paolo Manghi, Antonella Poggi,
  Giuseppe Serra, and Gianmaria Silvello, editors, {\em Proceedings of the 17th
  Italian Research Conference on Digital Libraries}, pages 117--125. CEUR
  Workshop Proceedings, 2021.

\bibitem{blackburn_etal}
Rob Blackburn, Thamirys Cabral, Ana Cardoso, Estelle Cheng, Pedro Costa, Paula
  Demain, Tom Demeranville, Dan Dineen, C~Dumitru, Padma Gopinath, and et~al.
\newblock Orcid public data file 2020.
\newblock https://orcid.figshare.com/articles/dataset/ORCID
  \_Public\_Data\_File\_2020/13066970/1, Oct 2020.

\bibitem{10.3390/publications4040030}
Josh Brown, Tom Demeranville, and Alice Meadows.
\newblock {Open Access in Context: Connecting Authors, Publications and
  Workflows Using ORCID Identifiers}.
\newblock {\em Publications}, 4(4):30, 2016.

\bibitem{clark_2020}
Rosa Clark.
\newblock Metadata deposit schema - crossref.
\newblock
  https://www.crossref.org/documentation/content-registration/metadata-deposit-schema/,
  Apr 2020.

\bibitem{clark_2021}
Rosa Clark.
\newblock New public data file: 120+ million metadata records - crossref, Jan
  2021.

\bibitem{Cohen2015}
Adam Cohen.
\newblock fuzzywuzzy.
\newblock \url{https://github.com/seatgeek/fuzzywuzzy}, 2015.

\bibitem{dasler_robin_2017_841777}
Robin Dasler, Adèniké Deane-Pratt, Artemis Lavasa, Laura Rueda, and Sünje
  Dallmeier-Tiessen.
\newblock {Study of ORCID Adoption Across Disciplines and Locations}.
\newblock December 2017.

\bibitem{group_2019}
ORCID Funder~Working Group.
\newblock Orcid and grant dois: Engaging the community to ensure openness and
  transparency of funding information.
\newblock https://orcid.figshare.com/articles/online\_resource/ORCID
  \_and\_Grant\_DOIs\_Engaging\_the\_Community\_to\_Ensure
  \_Openness\_and\_Transparency\_of\_Funding\_Information/9105101/1, Sep 2019.

\bibitem{10.1087/20120404}
Laurel~L. Haak, Martin Fenner, Laura Paglione, Ed~Pentz, and Howard Ratner.
\newblock {ORCID: a system to uniquely identify researchers}.
\newblock {\em Learned Publishing}, 25(4):259--264, 2012.

\bibitem{10.3389/frma.2021.656233}
Daniel~W. Hook and Simon~J. Porter.
\newblock {Scaling Scientometrics: Dimensions on Google BigQuery as an
  Infrastructure for Large-Scale Analysis}.
\newblock {\em Frontiers in Research Metrics and Analytics}, 6:656233, 2021.

\bibitem{10.3389/frma.2018.00023}
Daniel~W. Hook, Simon~J. Porter, and Christian Herzog.
\newblock {Dimensions: Building Context for Search and Evaluation}.
\newblock {\em Frontiers in Research Metrics and Analytics}, 3:23, 2018.

\bibitem{meadows_haak_2017}
Alice Meadows and Laurel Haak.
\newblock Orcid open letter - one year on report.
\newblock Apr 2017.

\bibitem{mejias_2020}
Gabriela Mejias.
\newblock Collect \& connect - improved and updated!
\newblock {\em ORCID}, Dec 2020.

\bibitem{Note1}
The idea of a centralised identity and authentication mechanism for academia is
  an alluring one. However, the idea that, at the current time, publishers,
  funders and academic institutions would all make themselves reliant on a
  centralised third-party is difficult to imagine. Furthermore, we live in an
  era where the direction of movement in technology is toward the
  decentralisation of trust or, more specifically, the distribution of trust
  across networks. Hence, it seems unlikely that centralisation in this context
  would be a wise structural choice at this time.

\bibitem{ARC_FOR}
Australian~Bureau of~Statistics.
\newblock Australian and new zealand standard research classification (anzsrc),
  Jun 2020.

\bibitem{orcid_open_letter}
ORCID.
\newblock Orcid in publications, Jan 2016.

\bibitem{orcid_2020_Aug}
ORCID.
\newblock Funders' orcid policies.
\newblock https://info.orcid.org/funders-orcid-policies/, Aug 2020.

\bibitem{orcid_2021_Apr}
ORCID.
\newblock Internal orcid weekly statistic report, Aug 2021.

\bibitem{orcid_2021}
ORCID.
\newblock Orcid record schema.
\newblock https://info.orcid.org/documentation/integration-guide/orcid-record/,
  Feb 2021.

\bibitem{science_porter_2016}
Simon Porter.
\newblock Digital science white paper: A new ‘research data mechanics’.
\newblock Aug 2016.

\bibitem{porter_2021}
Simon Porter.
\newblock Bringing narrative to research collaboration networks in 3d - digital
  science.
\newblock
  https://www.digital-science.com/blog/2021/05/3d-research-collaboration-networks/,
  2021.

\bibitem{orcid_finland}
Hanna-Mari Puuska.
\newblock Orcid in publications.
\newblock
  https://info.orcid.org/the-new-finnish-research-information-hub-provides-a-comprehensive-view-of-finnish-research/,
  Jun 2020.

\bibitem{ukri_2021}
UKRI.
\newblock Ukri open access policy.
\newblock
  https://www.ukri.org/wp-content/uploads/2021/08/UKRI-060821-UKRIOpenAccessPolicy-FINAL.pdf,
  Aug 2021.

\bibitem{UNESCO_2021}
UNESCO.
\newblock {\em UNESCO Science Report: the race against time for smarter
  development}.
\newblock UNESCO, 2021.

\bibitem{10.1162/qss_a_00112}
Martijn Visser, Nees Jan~van Eck, and Ludo Waltman.
\newblock {Large-scale comparison of bibliographic data sources: Scopus, Web of
  Science, Dimensions, Crossref, and Microsoft Academic}.
\newblock {\em Quantitative Science Studies}, 2(1):1--22, 2021.

\bibitem{10.1186/s13321-019-0365-4}
Egon Willighagen, Nina Jeliazkova, and Rajarshi Guha.
\newblock {Journal of Cheminformatics, ORCID, and GitHub}.
\newblock {\em Journal of Cheminformatics}, 11(1):44, 2019.

\end{thebibliography}

\end{document}